\documentclass[10pt,preprint,twocolumn]{aastex631}

\newcommand{\roma}[1]{\uppercase\expandafter{\romannumeral#1}}
\newcommand{\speed}[1]{#1 km~s${}^{-1}$}

\newcommand{\acc}[1]{#1 km~s${}^{-2}$}
\newcommand{\nfig}[1]{Figure~\ref{#1}}

\newcommand{\chenpengfei}{Rev. Geophys. Planet. Phys.}
\usepackage{amsmath}
\usepackage{subfigure}
\usepackage{hyperref}%

 \usepackage{url}
\usepackage{graphicx}
\usepackage{natbib}
\usepackage{amssymb,txfonts}
\usepackage{multirow}
\usepackage{array}
\usepackage{sidecap}
\usepackage{hyperref}
\usepackage{epstopdf}
\usepackage{footnote}
\usepackage{tabularx}
\usepackage{booktabs}

\usepackage{CJK}
\hypersetup{
 bookmarks=true,  
 unicode=false,  
 pdftoolbar=true, 
 pdfmenubar=true, 
 pdffitwindow=false, 
 pdfstartview={FitH}, 
 pdftitle={My title}, 
 pdfauthor={Author}, 
 pdfsubject={Subject}, 
 pdfcreator={Creator}, 
 pdfproducer={Producer}, 
 pdfkeywords={keyword1} {key2} {key3}, 
 pdfnewwindow=true, 
 colorlinks=true, 
 linkcolor=blue,  
 citecolor=blue,  
 filecolor=blue
 urlcolor=blue
 }

\shorttitle{Broad and Bi-directional narrow quasi-periodic fast-propagating wave trains associated with a filament-driven halo CME on 2023 April 21}
\shortauthors{Zhou et al.}

\newcommand{\xp}[1]{{\color{black}{#1}}}
\newcommand{\zxp}[1]{{{#1}}}

\begin{document}
\begin{CJK*}{UTF8}{gbsn}
\title{Broad and Bi-directional narrow quasi-periodic fast-propagating wave trains associated with a filament-driven halo CME on 2023 April 21}

\correspondingauthor{Xinping Zhou }
\email{xpzhou@sicnu.edu.cn}
\correspondingauthor{Yuandeng Shen }
\email{ydshen@ynao.ac.cn}

\author[0000-0001-9374-4380]{Xinping Zhou }
\affiliation{ College of Physics and Electronic Engineering, Sichuan Normal University, Chengdu 610068, People's Republic of China}
\affiliation{State Key Laboratory of Space Weather, Chinese Academy of Sciences, Beijing 100190}

\author[0000-0001-9493-4418]{Yuandeng Shen }
\affiliation{Yunnan Observatories, Chinese Academy of Sciences, Kunming 650216, People's Republic of China}

\author{Yihua Yan}
\affiliation{State Key Laboratory of Solar Activity and Space Weather, National Space Science Center of Chinese Academy of Sciences, Beijing 100084, China}
\affiliation{School of Astronomy and Space Sciences, University of Chinese Academy of Sciences, Beijing 100049, China}

\author[0000-0002-1134-4023]{Ke Yu}
\affiliation{ College of Physics and Electronic Engineering, Sichuan Normal University, Chengdu 610068, People's Republic of China}

\author[0000-0001-8318-8747]{Zhining Qu}
\affiliation{ College of Physics and Electronic Engineering, Sichuan Normal University, Chengdu 610068, People's Republic of China}

\author[0000-0001-9540-5235]{ Ahmed Ahmed Ibrahim}
\affiliation{Department of Physics and Astronomy, College of Science, King Saud University, P.O. Box 2455, 11451 Riyadh, Saudi Arabia}

\author[0000-0003-0880-9616]{Zehao Tang}
\affiliation{Yunnan Observatories, Chinese Academy of Sciences, Kunming 650216, People's Republic of China}

\author[0009-0005-5300-769X]{ Chengrui Zhou}
\affiliation{Yunnan Observatories, Chinese Academy of Sciences, Kunming 650216, People's Republic of China}

\author[0009-0005-5300-769X]{ Song tan}
\affiliation{Leibniz-Institut f{\"u}r Astrophysik Potsdam (AIP), An der Sternwarte 1614482 Potsdam, Germany}

\author[0000-0002-1190-0173]{Ye Qiu}
\affiliation{School of Astronomy and Space Science, Nanjing University, Nanjing 210023, China}

\author{ Hongfei Liang}
\affiliation{ Department of Physics, Yunnan Normal University, Kunming 650500, People's Republic of China}

\begin{abstract}
This paper presents three distinct wave trains \xp{occurred on 2023 April 21}: a broad quasi-periodic fast-propagating (QFP) wave train and \xp{bi-directional} narrow QFP wave trains. The broad QFP wave train expands outward in a circular wavefront, while \xp{bi-directional} narrow QFP wave trains propagate in the northward and southward directions, respectively.
The concurrent presence of the wave trains offers a remarkable opportunity to investigate their respective triggering mechanisms. Measurement shows that the speed of the broad QFP wave train is in the range of 300-\speed{1100} in different propagating directions. There is a significant difference in the speed of the \xp{bi-directional} narrow QFP wave trains: the southward propagation achieves \speed{1400}, while the northward propagation only reaches about \speed{550} accompanied by a deceleration of about 1-\acc{2}. Using the wavelet analysis, we find that the periodicity of the propagating wave trains in the southward and northward directions closely matches the quasi-periodic pulsations (QPPs) exhibited by the flares. Based on these results, the narrow QFP wave trains were most likely excited by the intermittent energy release in the accompanying flare. \xp{In contrast, the broad QFP wave train had a tight relationship with the \zxp{erupting} filament, probably \zxp{attributed} to the unwinding motion of the erupting filament, or the leakage of the fast sausage wave train inside the filament body.}

\end{abstract}


\section{Introduction} \label{sec:intro}

\zxp{In the last quarter century}, global extreme ultraviolet (EUV) waves have been observed in the solar corona \citep{1998GeoRL..25.2465T}. 
They are spectacular perturbations in the highly ionized and magnetized coronal plasma \xp{and have a tight relationship with the corona heating \citep{2020SSRv..216..140V} corona hole (CH) heating \citep{2005SSRv..120...67O} and the acceleration and heating of the solar wind \citep{1998JGR...10323677O,2010LRSP....7....4O}}. \xp{After a decade-long debate, their physical nature has \zxp{finally been identified}, that is, a fast-mode magnetosonic wave driven by the lateral expansion of the accompanying coronal mass ejections (CMEs), from the modeling \citep{2001JGR...10625089W} and the observation \citep{2018ApJ...864L..24L,2020ApJ...905..150Z}. Its true wave features have been observed, such as reflection, transmission, when they interacted with active regions and CHs \citep{2009ApJ...691L.123G,2010ApJ...723L..53L,2012ApJ...756..143O,2012ApJ...746...13L,2013ApJ...773L..33S,2021A&A...651L..14M}. Notably, \cite{2002ApJ...574..440O} and \cite{2010ApJ...713.1008S} first presented the 3D MHD simulation of the interaction of the large-scale coronal wave with the active region and CH, respectively, where the wave undergoes strong reflection and refraction, in agreement with the observation reported by \cite{1999SoPh..190..467W} and \cite{2009ApJ...691L.123G}.}
For a detailed research process on EUV wave, please refer to the review articles \citep{2009SSRv..149..325W,2014SoPh..289.3233L,2015LRSP...12....3W,2016GMS...216..381C,2017SoPh..292....7L,2022chenpengfei}. In the Solar Dynamics Observatory \citep[SDO;][]{2012SoPh..275....3P} era, its Atmosphere Imaging Assembly \citep[AIA;][]{2012SoPh..275...17L}, with high spatial and temporal resolution imaging capabilities, captured many details of the EUV waves, such as the narrow and broad quasi-periodic fast-propagating (QFP) wave trains with multiple wavefronts \citep[see the review by][]{2014SoPh..289.3233L,2022SoPh..297...20S}.

The narrow QFP wave train was first reported by \cite{2011ApJ...736L..13L} utilizing the AIA imaging data, \xp{ which were identified as a fast magnetosonic wave train by \cite{2011ApJ...740L..33O} using the 3D MHD model.} These wave trains have some significant features different from the classical large-scale EUV wave that propagates along the solar surface, which is dominated by the vertical magnetic field line, such as permanently confined propagating along the closed- or open-loops system. \xp{In particular, \cite{2018ApJ...860...54O} first studied the counter-propagating narrow QFP wave trains along the closed trans-equatorial coronal loops system. Their study showed that these wave trains interacted in the middle of the loop system, resulting in the excitation of trapped kink-mode and slow-mode MHD waves. These findings suggest that the counter-propagating QFP wave trains within closed coronal loops can generate a turbulent cascade capable of carrying substantial energy for heating the corona in low-corona magnetic structures.} Generally, it was found that the narrow QFP wave trains appear similar periodicities with the quasi-periodic pulsations (QPP) \citep{2011ApJ...736L..13L,2012ApJ...753...53S,2013A&A...554A.144Y,2018ApJ...868L..33L,2022ApJ...926L..39D,2022ApJ...941...59Z} of the accompanying flare emissions commonly seen from radio to hard X-rays \citep{2009SSRv..149..119N,2015ApJ...807...72L,2016ApJ...832...65Z,2020ApJ...893....7L,2021ApJ...921..179L,zhang2024}. This result aligns with previous studies indicating that flares can produce non-thermal electrons, which are accountable for microwave and X-ray emissions. Additionally, flares can induce MHD oscillation. The correlation between QPP in microwave and X-ray emissions and MHD waves has been extensively examined \citep[see review by][and references therein]{2009SSRv..149..119N}. However, other viewpoints favor that the dispersion causes the narrow QFP wave trains. The inhomogeneous density and magnetic field of the loop system act as a waveguide. This special structure provides a condition for dispersion of a fast magnetoacoustic wave with broadband when they propagate confined in the loop systems \citep{1993SoPh..144..255M,2017ApJ...847L..21P}. 

\xp{The first unambiguous imaging of the broad QFP wave train was noted by \citep{2012ApJ...753...52L}}. This type of QFP wave train has many characteristics that are similar to the classical large-scale EUV wave, such as they will display refraction and transmission when they interact with coronal structures such as cavities \citep{2012ApJ...753...52L}, filament \citep{2019ApJ...873...22S} and CH \citep{2022ApJ...930L...5Z,2024NatCo..15.3281Z}. Notably, \cite{2022A&A...659A.164Z} first reported and certified the \emph{total reflection} of a broad wave train after collision with a south CH, which enriches the wave feature of the coronal waves. Although the broad QFP wave train shares many common appearance features, such as the speeds, intensity, and some true wave characteristics as mentioned above, it raises a new issue that its driving mechanism is hard to explain using the CME expansion mechanism, where only one wavefront due to the compress of the expansion CME \citep{2020MNRAS.493.4816M}. Furthermore, \cite{2021SoPh..296..169Z} reported a broad QFP wave train first propagated inside the CME bubble and then caught up with the CME front, eventually running in front of the CME front. \xp{\cite{2012ApJ...753...52L} discovered broad QFP wave trains within a single pulse EUV wave propagating ahead of the lateral CME front. The dominant 2-minute periodicity of the wave train matches the X-ray flare pulsation. Therefore, they considered the flare to have a tight connection with the \zxp{excitation} of the broad QFP wave train, supported by \cite{2021ApJ...911L...8W} and \cite{2022ApJ...930L...5Z}. Recently, \cite{2024ApJ...962...42H} proposed a new excitation mechanism for these wave trains. Their simulations show that perturbations propagating outward from the internal magnetic rope through its surface generate wave trains similar to those reported by \cite{2012ApJ...753...52L} running ahead of the CME's flanks. On the other hand, \cite{2019ApJ...873...22S} and \cite{2024arXiv240106661Z} argue that the broad QFP wave train is causally connected to the successive expansion and unwinding of filament threads, as they share a common dominant period.} These new details imply that the \xp{excitation} mechanism of the EUV wave is still an open question.

The detailed parameters of these two QFP wave trains, such as speeds, periods, and intensity amplitude, can be found in the recent statistics conducted by \cite{2022SoPh..297...20S} that were obtained from the published articles. In the present work, we investigate the trigger mechanism of multiple wave trains and the parameters between them. The data from the H$\alpha$ Imaging Spectrograph \citep[HIS;][]{2022SCPMA..6589605L} on board the Chinese H$\alpha$ Solar Explorer \citep[CHASE;][]{2022SCPMA..6589602L}, the Solar Upper Transition Region Imager \citep[SUTRI;][]{2023RAA....23f5014B} on board the Space Advanced Technology demonstration satellite (SATech-01), the ground-based New Vacuum Solar Telescope \cite[NVST;][]{2014RAA....14..705L,2016NewA...49....8X,2020ScChE..63.1656Y}, SDO/AIA and other instrument are analyzed, to show the erupting filament as well as the coronal magnetic environment. The primary analysis results are presented in Section \ref{se:results}. The discussion and conclusion are given in Section \ref{se:discussion}.

 \begin{figure}
	\centering
	\includegraphics[width=1.\linewidth]{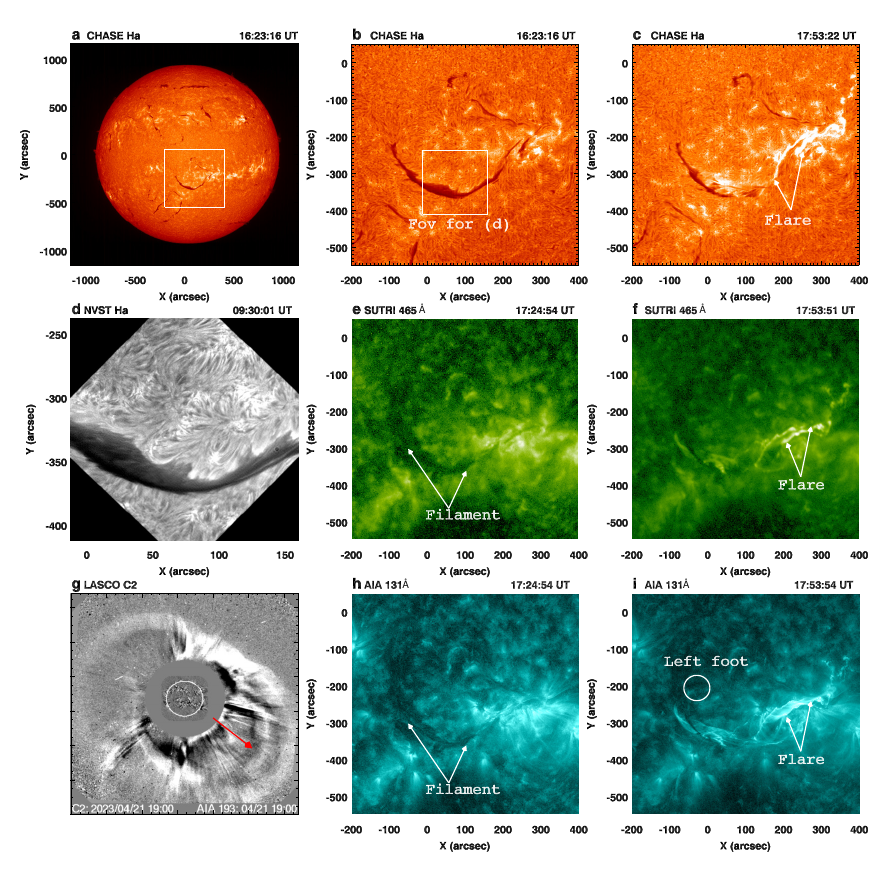}
	\caption{Panels (a)-(c) show the location and the erupting process of the filament of interest in the HIS/CHASE H$\alpha$ center (6562.8 \AA) images. Panels (e)-(f) and (h)-(i) are the snapshots to show the filament in SUTRI 465 \AA\ and AIA/SDO 131 \AA\ images. Panel (d) displays the filament of interest in 9:30 UT observed by the NVST in H$\alpha$ waveband. Panel (g) shows the halo CME observed by the C2/LASCO. The white arrows point to the filament of interest and the flare ribbon, while the red arrow in panel (g) indicates the inner fronts and the expulsion direction observed in the low corona. Panels (e)-(f) and (h)-(i) are the enlarged region marked with the white box in panel (a), while panel (d) is the enlarged region labeled with a white box in panel (b). 
		\label{fig:overview}}
\end{figure}

\section{Observations and Results}

\label{se:results}
The filament eruption event of interest occurred on 2023 April 21 and was well observed by the CHASE, SUTRI, NVST, and SDO. The CHASE, launched on 2021 October 14, is the first solar space mission of the China National Space Administration (CNSA). The HIS instrument onboard can provide spectroscopic observations of the Sun by scanning the full solar disk in both H$\alpha$ (6559.7-6565.9 \AA) and Fe I (6567.8-6570.6 \AA) wavebands with a spectral resolution of 0.024 \AA\ per pixel, a spatial resolution of 0$\farcs$52 per pixel and a temporal resolution of 60 s. The HIS/CHASE data used here have been calibrated \citep{2022SCPMA..6589603Q}. SUTRI image the upper transition region fo the Sun \citep{2017RAA....17..110T}, uses the Ne V\footnotesize{II}\normalsize\ 465 \AA\, which forms at a temperature of about 0.5 MK. Its field of view is 41$\farcs$6 $\times$ 41$\farcs$6, with a spatial and temporal resolution of about 8$\farcs$0 and 30 seconds, respectively. The NVST \citep{2014RAA....14..705L}, located in the Fuxian Solar Observatory of the Yunnan Observatories, Chinese Academy of Sciences (CAS), aims to observe the delicate structures of the photosphere and the chromosphere. Its H$\alpha$ line-center image, acquired at 6562.8 \AA\ with a spatial and temporal resolution 0\farcs3 and about 12 seconds, respectively, is employed to demonstrate the filament. 
The AIA images' time cadence and pixel size are 12 seconds and 0$\farcs$6. We mainly utilize the image observations of 171 \AA, 193 \AA, and 211 \AA\ EUV passbands to study the wave trains. All the AIA data used here are calibrated with the standard procedure \texttt{aia\_prep.pro} available in the \texttt{SolarSoftWare} package provided by instrument team.

The large-scale C-shape filament was located on the southwest part of the solar disk, as shown in \nfig{fig:overview} (a). \nfig{fig:overview} (b) and (c) show two snapshots of the eruption process in CHASE H$\alpha$ observation, with the zoom-in region of the panel (a) highlighting a distinct filament. From the viewpoint of NVST, one can find that the filament remained quiet station (see \nfig{fig:overview} (d) captured at 09:30 UT) for several hours before the eruption, until 17:25 UT, its west partial generally lifted, and eventually erupted entirely at around 18:00 UT (see \nfig{fig:overview} (c)). The complete eruption process of the filament was also observed in the EUV passbands, as shown in the \nfig{fig:overview} (e)-(f) and (h)-(i) that were taken from the SUTRI and AIA, respectively. The eruption coincided with a GOES M1.7 flare, a hale CME with a two-front morphology (see \nfig{fig:overview} panel (g)) \citep{2019ApJ...870...15L,2022A&A...665A..51S} and a series of QFP wave trains. \xp{The eruption process will be discussed in the following paper, using CHASE data, as its full-disk \zxp{spectroscopic imaging has} the advantage of investigating the 3D kinematics of solar filament eruptions \citep{2024ApJ...961L..30Q}}. Here, we mainly study the exciting mechanism and the parameters of the QFP wave trains.

\begin{figure}
	\centering
	\includegraphics[width=1.\linewidth]{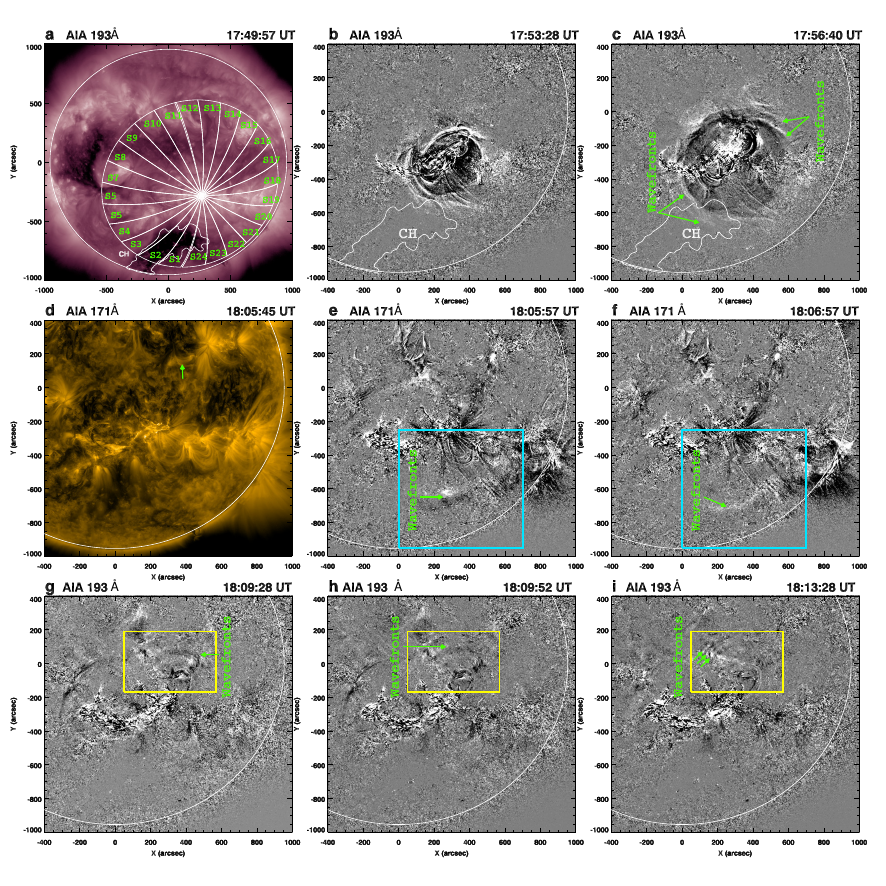}
	\caption{Panels (a) and (d) are the AIA/SDO 211 \AA\ and 171 \AA\ direct images to show the initial coronal condition of the eruption source region. The closed region in panel (a), denoted by ``CH'', represents the southern CH, whose boundary is also projected onto panels (b) and (c). The sectors, labeled with S1-S24 and centered on the flare, in panels (a) are used to obtain time-distance stack plots in \nfig{fig:tdp}. Panels (b) and (c) are the snapshots of the broad QFP wave train at different times using the AIA 193 \AA\ running difference images. The green arrows marked the wavefronts' positions in the quiet-Sun region and inside the CH area. Panels (e)-(f) and (g)-(i) display the evolution of the southward and northward propagating wave trains, respectively. The green arrows mark the location of the wavefronts, while the blue and yellow rectangles show the field view of the images in \nfig{fig:south} and \nfig{fig:north}, respectively. The white curves mark the disk limb (the same in \nfig{fig:south}). The animation covers 17:20 UT-18:30 UT with a 12 seconds cadence. The animation duration is 12 seconds. (An animation of this figure is available.) 
		\label{fig:evolution}}
\end{figure}

From the overview of SDO, one can find that a high-latitude CH can be found located to the southeast of the source (see \nfig{fig:evolution} (a)). The outline of the CH is also projected onto the \nfig{fig:evolution} (b) and (c). To visualize the wavefront signal, we utilize the running-difference images of AIA to study its evolution and kinematics. It should be pointed out that the running-difference images are created by forward subtracting images with a time interval of 12 seconds. In the running-difference images, the moving wavefronts are seen as a bright arc-shaped contour followed by a black one. Unlike the loops-oscillation-associated and jet-associated wavefronts that propagated mainly limited in the front of the jet \citep{2023ApJ...953..171H,2022ApJ...931..162Z} and loops \citep{2018ApJ...860L...8S,2022ApJ...939L..18S,2023A&A...674A.167L}, the broad QFP wave with multiple circular-shaped wavefront (see the green arrows in \nfig{fig:evolution} (b) and (c)) concentric with the eruption kernel propagated outward along the solar surface at about 17:48 UT. About 12 minutes later, at 18:00 UT, a series of narrow wavefronts appeared, propagated mainly in the southwest direction (\nfig{fig:evolution} (e) and (f)). At 18:07 UT, another series of narrow wavefronts appeared and propagated in the direction of the northeast (see the green arrows in \nfig{fig:evolution} (g)-(i)). Comparing the snapshots of these wave trains, we can find that each has distinctive features. \zxp{In addition, from the animation of \nfig{fig:evolution}, one can see that the broad wave train signals of 193 \AA\ and 211 \AA\ are much more precise than those in 171 \AA. As discussed by \cite{2012ApJ...753...52L}, this pattern signifies the rapid heating of plasma, elevating temperatures from 0.8 to 2.0 MK, followed by subsequent cooling, likely due to the plasma undergoing adiabatic compression and subsequent rarefaction. On the contrary, as the southward propagating narrow wave train observed in the present event, they are always detected in AIA 171 \AA\ and occasionally in the 193 \AA\ and 211 \AA\, indicating the narrow temperature range. According to the explanation given by \cite{2016AIPC.1720d0010L}, the narrow temperature range is possibly due to two reasons: 1. owing to the low-intensity amplitude of the narrow wave train, which is reasonable that the temperature of the wave-hosting plasma is close to the AIA 171 \AA\ channel's peak-response temperature. In addition, the low-intensity amplitude of narrow wave trains makes it hard for them to cause significant temperature departures, unlike the broad wave trains, which can cause large intensity amplitude. 2. The varying sensitivities of the detectors in different AIA channels. The AIA 171 \AA\ channel, in particular, exhibits a significantly higher photon-response efficiency compared to the other channels, surpassing them by at least one order of magnitude. As a result, it displays a heightened sensitivity to even slight intensity variations. However, the northward propagating narrow wave train is seen better in 193 \AA\ and 211 \AA\ channels, whereas it appears fainter in the 171 \AA\ channel. According to the explanation above, this abnormal phenomenon indicates that the narrow wave train occasionally has a high-intensity amplitude, making it more evident in 193 \AA\ and 211 \AA. }

\subsection{Broad QFP wave train}
\begin{figure}
	\centering
	\includegraphics[width=1.\linewidth]{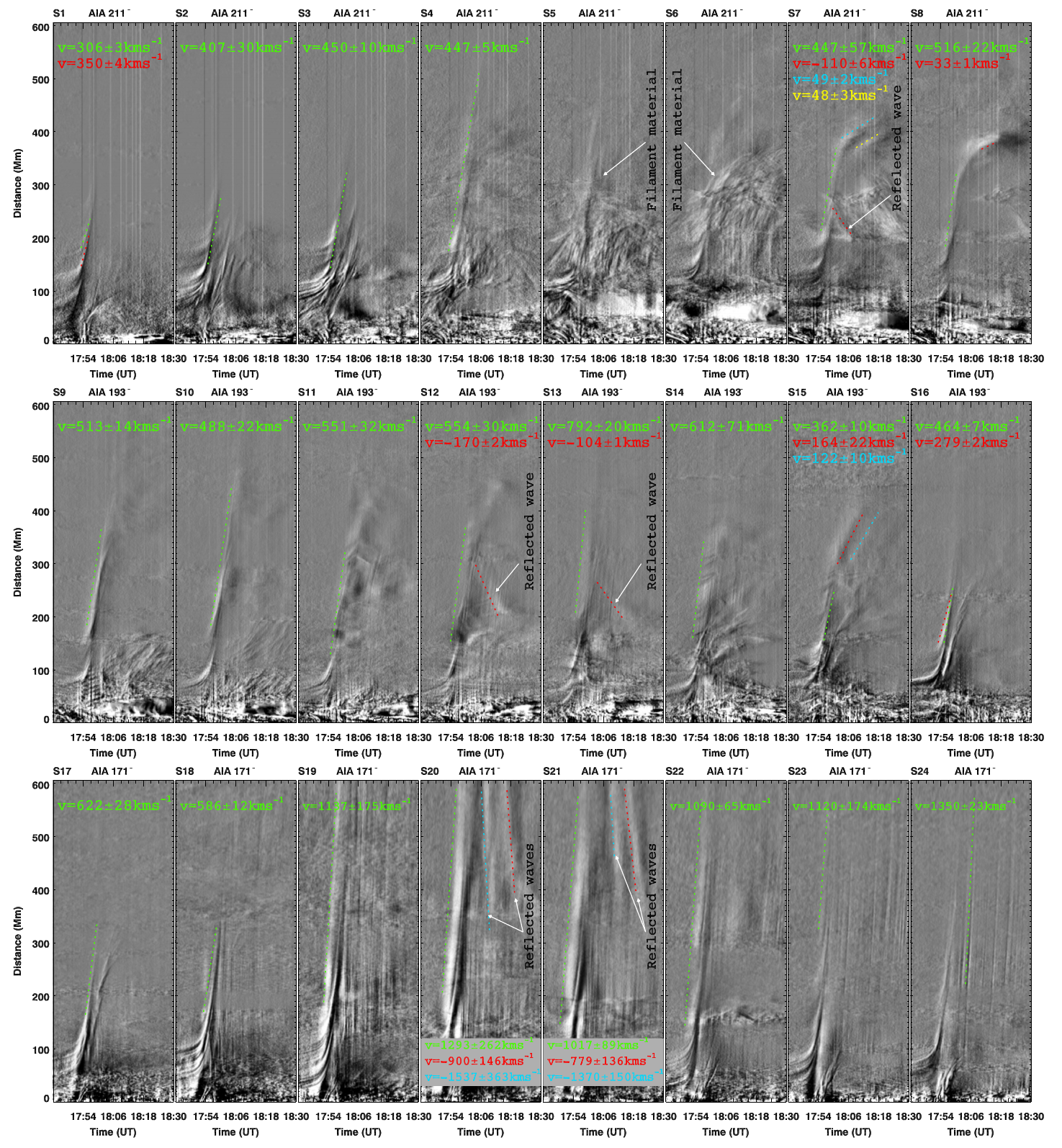}
	\caption{ Time-distance stack plots reconstructed along the sectors-shaped slices S1-S24 in \nfig{fig:evolution}, using the AIA 211 \AA\ (S1-S8), 193 \AA\ (S9-S16) and 171 \AA\ (S17-S24) running difference images. The ridges' slope represents the moving material's speeds, and their speeds obtained by linear fitting are listed in each panel with corresponding colored fonts.
		\label{fig:tdp}}
\end{figure}

To study the kinematic of the broad wave train, we made the time-distance stack plots (TDSPs) along 24 sectors S1-S24, each spanning 15$^\circ$ (\nfig{fig:evolution} (a)) and centering on the eruption center. Although the broad QFP wave train has been captured by the AIA 211 \AA, 193 \AA, and 171 \AA\ in different wavebands, they have some other characteristics. So, we respectively selected the AIA 211 \AA, 193 \AA, and 171 \AA\ running-difference images along sectors S1-S8, S9-S16, and S17-S24 to show the evolution of the wave train in TDSPs, as shown in \nfig{fig:tdp}. As the source is near the south CH, the wavefront can be traced to propagate about a distance of 300 Mm with a speed of 300-\speed{450} and eventually disappeared in the region of CH in the directions of S1, S2 and S3, which is different with the appearances, such as the reflection and transmission, reported in previous works \citep{2022A&A...659A.164Z,2022ApJ...930L...5Z}. Due to the affection of the erupting filament, the wave signal is hard to discern in the directions of S5 and S6 (see the arrows in S5 and S6). Interestingly, a partial wavefront appeared reflection with a speed of \speed{110} below the initial speed of \speed{447} when it interacted with the left foot of the filament marked with a white circle in \nfig{fig:overview} (see \nfig{fig:tdp} (S7)). Immediately afterward, the wavefront that continues to propagate forward shows an apparent deceleration and approaches a speed of about 30-\speed{50} after its impact with a magnetic structure near the CH2. Notably, in the direction of S7, the wave exhibits two wavefronts, while these multiple wavefronts feature is difficult to distinguish before the deceleration in this direction.

The wave was also reflected with a speed of 100-\speed{170} in the directions of S12-S13 when it interacted with a loop system marked with a green arrow in \nfig{fig:evolution} (d). When the wave approached a active region on its path in the direction of S15, it decelerated to 122-\speed{164} and exhibited two clear wavefronts (see \nfig{fig:tdp} S15). As shown in \nfig{fig:evolution} (d), multiple active regions were east of the source. Thus, the speeds (1000-\speed{1200}) in the S19, S20, and S21 directions were significantly higher than in other directions. Meanwhile, after interacting with the active region, the wave also showed a more substantial reflection with an speed of 800-\speed{900}. \xp{As mentioned above, the speeds of reflected waves are slower than that of the incident wave, which is still puzzling. One possibility is that the magnetic and plasma conditions are modified by the passage of the shock or CME \citep{2004JGRA..10912105G}, where the reflected wave will travel. Notably, the reflected wave with a speed of about \speed{100} even well below the sound speed ($c_{s}= 150-$\speed{210}) in the directions of S7 and S13. Since these reflected waves occurred far from the center of the solar disk, we speculate that the projection effect may have resulted in the reduced apparent speed of the reflected wave and thus contributed to this discrepancy. On the other hand, as shown in \nfig{fig:tdp} S7, the filament threads significantly interfered with the reflection signal in this direction, \zxp{leading to more significant uncertainty}. In addition, the reflected wave signals are relatively weak, which makes it difficult to determine the real wave signal feature in the time-distance plots. }

\subsection{Narrow QFP wave trains}

Following the broad QFP wave train, \xp{a bi-directional} narrow QPF wave trains propagating southward and northward were excited consecutively. as shown in \nfig{fig:evolution}. Although the signal of these two narrow QFP wave trains has been identified in \nfig{fig:tdp} obtained from the sectors S1-S24, the propagating direction was not along the radial direction. Thus, we re-select the slices to study their kinematics. 

\begin{figure}
	\centering
	\includegraphics[width=1.\linewidth]{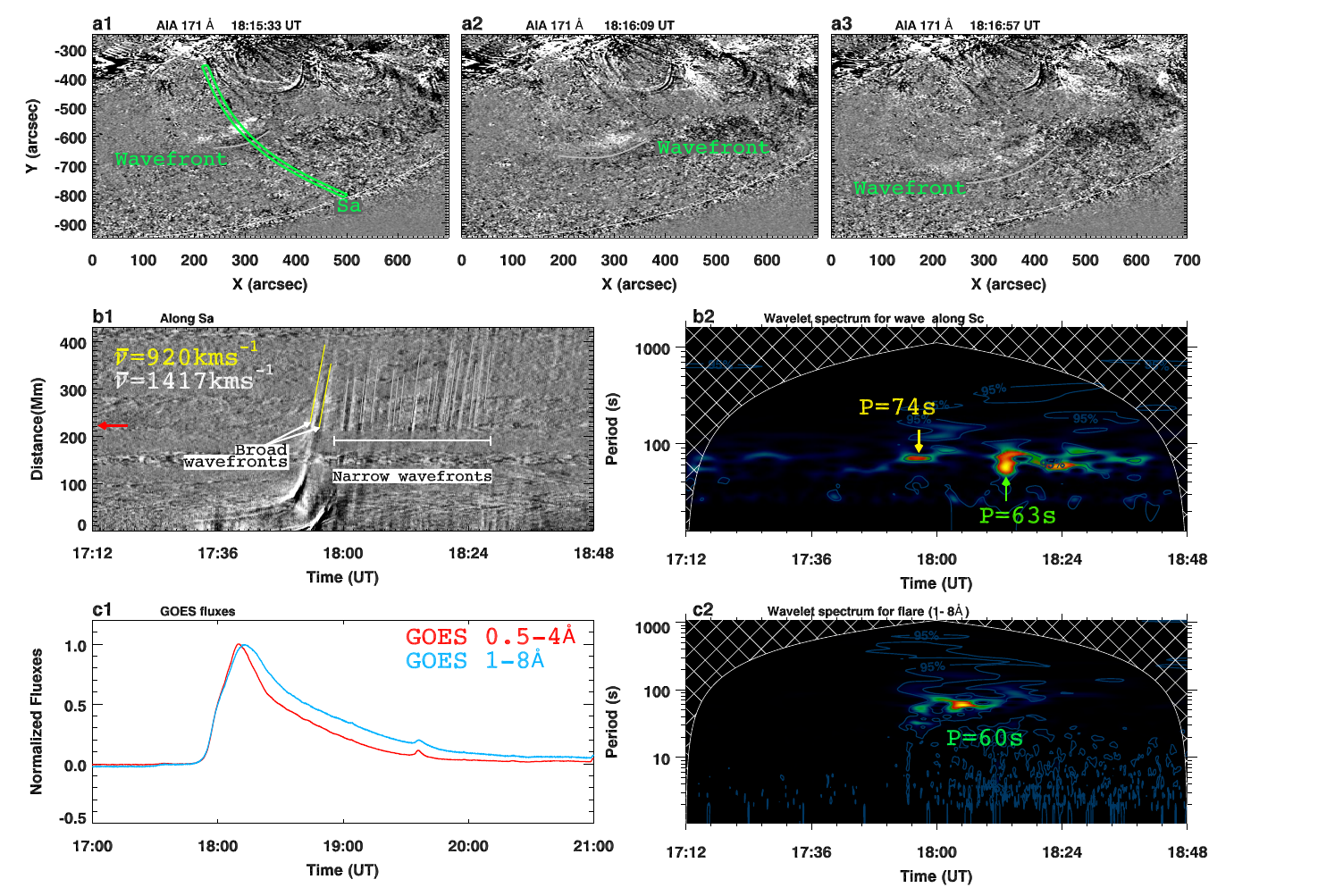}
	\caption{Panels (a1)-(a3), three snapshots, show the evolution of the southward propagating narrow QFP wave train using the AIA 171 \AA\ running difference images. The gray curves tracing the wavefronts are added to visualize the wave evolution (the same in \nfig{fig:north}), which is drawn by connecting a sequence of measurement points. The time-distance stack plot reconstructed along the slice labeled with ``Sa'' in panel (a1) is displayed in panel (b1). The fitting results of the broad and narrow QFP wavefronts are depicted with yellow and gray lines, respectively. Panel (b2) shows the wavelet spectrum of the detrended intensity profiles at the distance of 280 Mm, pointed by a red arrow in panel (b1). Panel (c1) displays the normalized GOSE 0.5-4.0 \AA\ and 1-8 \AA\ X-ray flux curves. The wavelet spectrum of the flare QPPs is shown in Panel (c2), using the GOSE 1-8 \AA\ detrended curves as an input. The results of wavelet analysis are listed in corresponding panels (b2) and (c2).
		\label{fig:south}}
\end{figure}

As shown in \nfig{fig:south} (a1)-(a2), the southward propagating narrow QFP wave train had an angular width of about 45$^\circ$ and bent to the southwest as propagating. We selected a slice marked with Sa in \nfig{fig:south} (a) to trance its evolution, and the corresponding TDSP is displayed in \nfig{fig:south} (b1). From panel (b1), the southward propagating narrow QFP wave train follows the broad QFP wave train. Meanwhile, the wavefronts did not appear to have significant deceleration. Using the linear fitting, we find the mean speed of the broad and narrow QFP wave trains were about \speed{920} and \speed{1417}, which are similar to that gotten along S22-S24 in \nfig{fig:tdp}. To explore the correlation between the wave train and the flare, we employ the wavelet analysis method \citep{1998BAMS...79...61T} to derive the periods of the wave train and the accompanying flare. For the wave trains, we first extracted the intensity 
profile at the distance of 220 Mm, marked with a red arrow in \nfig{fig:south} (b1), from the origin of coordinates. Then, by utilizing the detrended intensity profile as the input for wavelet analysis, the obtained wavelet spectrum is shown in the \nfig{fig:south} (b2). From the wavelet spectrum in \nfig{fig:south}, one can find that the broad and southward narrow wave train periods were 74s and 63s, respectively. For the flare, we use the soft X-ray flux recorded by GOES to investigate the fine structure of the flare pulsation (see \nfig{fig:south} (c1)). By using the derivative of the GOES Soft X-ray flux in the energy band of 1-8 \AA\ as the input, we get its dominant period is about 60s, as shown in the wavelet spectrum of \nfig{fig:south} (c2). The southward wave narrow wave train is similar to that of the flare. We speculate that the generation of the southward propagating narrow wave train might have a close relationship with the pulsed energy release in the accompanying flare.

 \begin{figure}
	\centering
	\includegraphics[width=1\linewidth]{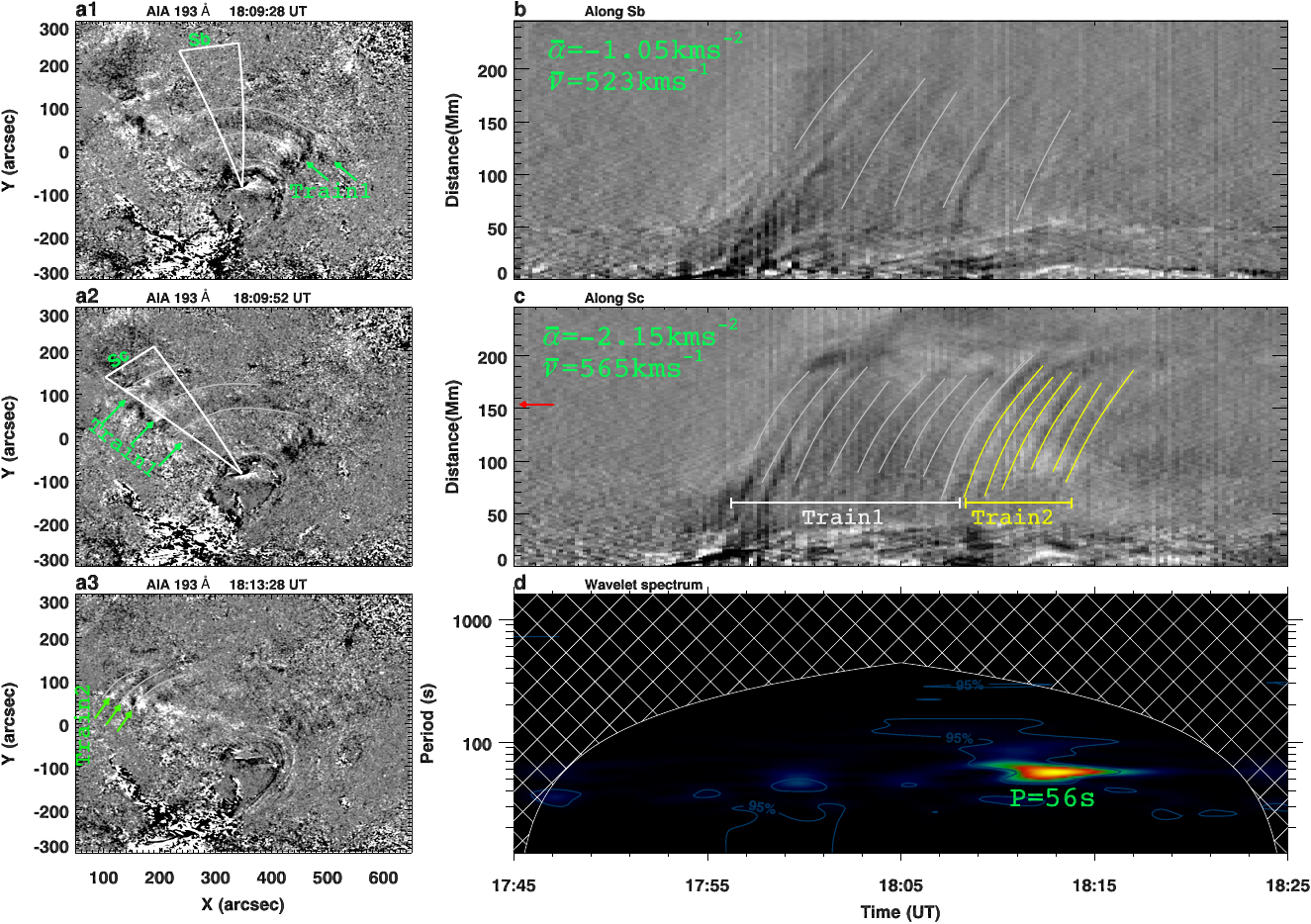}
	\caption{Panels (a1)-(a3) show the snapshots of the northward propagating narrow QFP wave train at three times using the AIA 193 \AA\ running difference images. The green arrows marked the wavefronts' positions. Panels (b) and (c) are the time-distance stack plots obtained along the sectors that marked ``Sb'' and ``Sc'' in panels (a1) and (a2), respectively. 
	}\label{fig:north}
\end{figure}

\nfig{fig:north} (a1)-(a3) show the enlarged field of view marked with a yellow box in \nfig{fig:evolution} (g)-(i), where one can find that the northward propagating narrow wave could be divided into two wave trains, Train1 and Train2. The wavelength of Train1 (see the green arrows in \nfig{fig:north} (a1) and (a2)) is significantly longer than that of Train2 (see the green arrows in \nfig{fig:north} (a3)). We select two sectors, labeled Sb and Sc in \nfig{fig:north} (a1) and (a2), to explore their kinematics, and the corresponding TDSPs are displayed in \nfig{fig:north} (b) and (c). Comparing the TDSPs in panels (b) and (c), we can identify that the signal of the wave train along the Sc is richer than that along Sb and the wave shows a significant deceleration. Using the quadratic fitting, we find that the mean speeds and the deceleration along these two paths were not significantly different, i.e., they are about \speed{550} and \acc{1.5}. Meanwhile, we can find that the period of Train2 is smaller than that of Train2. Similar to the method for analysis of the periodicity of the southward propagating wave train, by using the intensity profile at the distance of about 150 Mm (see the red arrow in \nfig{fig:north} (c)) as the input for the wavelet analysis, we obtain that the dominating period of Train2 is about 56s. However, we do not get the periodicity of Train1 in the wavelet spectrum due to its signal being too weak to be detected. According to the duration and the number of the wavefronts, the period of Train1 is calculated to be about the 70 s, which is also similar to the period of the QPP in the accompanying flare.

\section{Discussion and conclusions} 
\label{se:discussion}
Using the high-resolution spatial and temporal observations, we have presented a rare observation of a broad and \xp{bi-directional} narrow QFP wave trains consecutively excited during the flare on 21 April 2023. The eruption of the filament \xp{of interest} results in GOES M1.7 and halo CME formation. The first excited broad wave train with a circular-shaped wavefront propagated outward with speed in the range of 300-\speed{1100} in different directions. Notably, its speed was high to about \speed{1000} when it was across the active regions west of the source. The broad wave train appeared to reflect when interacting with the magnetic structures, making the multiple wavefronts more easily distinguished. The subsequently excited southward and northward narrow QFP wave trains have a smaller angular width (about 45$^\circ$). The southward narrow QPF wave train propagated with a mean speed of \speed{1400} and did not appear deceleration. On the contrary, the northward narrow QFP wave train exhibited stronger deceleration, and its speed (about \speed{550}) was slower than that of the northward narrow QFP wave train. Considering that the southward and northward narrow QFP wave trains share a common period of about 60s with the QPPs in the accompanying flare, we speculate that these two narrow QFP wave trains might have a stronger connection with the impulsive energy-releasing process in the accompanying flare. While the broad QFP wave train should has a causally connection with the erupting filament. 

Studying the coronal wave is essential for solar physics since it can provide insights into the perturbation's physical nature and its generation mechanism, \xp{and also for coronal seismology \citep{2005LRSP....2....3N}}. As the SDO/AIA was launched, new features have been captured, such as the characteristic circular-shaped multiple wavefronts and the narrow QFP wave trains reported here, which are differ from the classical large-scale EUV with a single wavefront. With the discovery of these new features, the origin of the coronal wave, especially for the large-scale wave, which has led to decades of debate, appears to be open. 

It is now widely recognized that the large-scale single EUV wave excitation and the CMEs are closely related. In this scenario, the piston (CME) can generate a shock wave ahead of the CME, which will freely propagate once it decouples from the CME. \xp{This scenario was first confirmed by STEREO observation reported by \cite{2009ApJ...700L.182P} and modeled using 3D MHD model by \cite{2010ApJ...713.1008S}}. In this excitation mode, the excited wave generally shows only one wavefront, certified in many observations \citep{2011ApJ...738..160M,2012ApJ...745L...5C}. Therefore, using this mechanism to explain the multiple wavefronts of the broad QFP wave train reported in this work is hard. The second view believes that the broad QFP wave train is driven by the pressure pulse caused by the intermittent energy release in the accompanying flare. In this scenario, the broad wave train often exhibits a dominant period that matches the pulsation of the accompanying flare.
Meanwhile, the beginning time of the broad QPF wave train usually lags behind flare QPPs onset time \citep{2022ApJ...930L...5Z} due to the excited wave needing time to evolve to be observed by the observation instruments \citep{2012ApJ...753...52L}. \xp{\cite{,2021ApJ...911L...8W} reproduced a dome-shaped broad QFP wave train propagating perpendicular to the magnetic field lines at a speed of approximately 550-\speed{700}. According to the author, this wave train was likely generated by the energy release associated with flare QPP.} In the present case, the flare QPPs initiation time is later than that of the broad QFP wave train. Thus, the broad QFP wave train studied here should probably have little to do with the flare.
Recently, \cite{2022ApJ...939L..18S} found that the broad QFP wave train might be driven by the successive stretching of the magnetic field lines during the eruption of the filament. In the observed case, the speed of the wave train is about three times the inner edge of the wave train, which is in agreement with the predictions in the model proposed by \cite{2002ApJ...572L..99C}. \xp{\cite{2024ApJ...962...42H} provide a new perspective for understanding the triggering of the broad QFP wave trains. They found that the leakage of the internal disturbance in the magnetic rope acting as a waveguide to form a multi-wavefront structure is a reasonable mechanism for generation wave trains. Their simulation result is consistent with the observation provided by \cite{2012ApJ...753...52L}. Interestingly, in certain exceptional cases, researchers find that the broad QFP wave trains could be driven by the unwinding of the filament threads \citep{2019ApJ...873...22S,2024arXiv240106661Z}, which is the \zxp{periodic energy release} of the magnetic twist stored in the filament into the outer corona \citep{1986SoPh..103..299S,1996ApJ...464.1016C,2011ApJ...735L..43S,2013RAA....13..253H,2017ApJ...851...67S,2021RSPSA.47700217S,2021ApJ...911...33C}. We speculate that the broad QFP wave in this study should have a close relationship with the \zxp{erupting} filament, driven by the unwinding motion of the erupting filament
or the leakage of the disturbance from the internal of the magnetic rope. On the other hand, the pulsed energy release caused by quasi-periodic magnetic reconnection can also lead to the generation of QFP waves. In the process of magnetic reconnection, the magnetic-field energy converts to the kinetic and thermal energy of the plasma and non-thermal high-energy particle energies, which is essential to launch pulsed energy release and, therefore, trigger the QFP wave train. For example, reconnection at X-type null points \citep{2012ApJ...749...30M, 2016ApJ...829L..33L,2022A&A...665A..51S,2022ApJ...936L..12W}, repetitive generation and coalescence of plasmoids \citep{2000A&A...360..715K}, or current sheet fluctuations induced by super-Alfv\'enic beams and associated Kelvin–Helmholtz instability nonlinear oscillations \citep{2006ApJ...644L.149O}. Many numerical simulations based on the magnetic reconnection successfully reproduce the broad QFP wave trains with physical parameters consistent with observation, such as morphology, speed, period, and intensity amplitude. For instance, \cite{2015ApJ...800..111Y} performed a numerical \zxp{study} base on the interchange reconnection and found that the waves were successively launched by the collision between the plasmoids and the field in the outflow region.  \cite{2021ApJ...909...45Y} found that the reconnection outflows can help the turbulence under the CME to develop and generate wave \zxp{trains}, indicating the turbulence \zxp{is an important factor in producing wave trains around CMEs}. These mechanisms should also contribute to generating wave trains in present events.}

The excited mechanism of the narrow QFP wave train is also an open question. Generally, the most likely relevant driven mechanisms are the energy release in magnetic reconnections process \citep{2011ApJ...736L..13L,2012ApJ...753...53S,2013A&A...554A.144Y} and the dispersive evolution \citep{1983Natur.305..688R, 2013A&A...560A..97P, 2014A&A...569A..12N, 2017ApJ...847L..21P, 2018MNRAS.477L...6S}. 
\xp{\cite{2011ApJ...736L..13L} first reported that the association between the flare pulsation detected in hard X-ray by RHESSI and the narrow QFP wave trains periodicities:} The narrow QFP wave trains exhibit a close physical relationship with the accompanying flares, always sharing similar periods and having a close temporal and spatial association \citep{2012ApJ...753...53S, 2013SoPh..288..585S, 2018ApJ...853....1S}. In some instances, multiple narrow wave trains with different properties are excited one after another, and each wave train is accompanied by an energy burst \citep{2013A&A...554A.144Y, 2020ApJ...889..139M,2022A&A...659A.164Z}, suggesting a strong connection between the flares and the narrow QFP wave trains. Observation in this study supports this suggestion since the narrow QFP wave trains have a similar period to the pulsation period of the accompanying flare. Meanwhile, as the energy releasing of the flare mainly dominates during the Train2 of the northward narrow wave trains (see \nfig{fig:south} (c2)), resulting in the intensity of the Train2 is significantly stronger than that of the Train1 (see \nfig{fig:north}) (c)).

 \xp{Based on the analyses presented above, we can infer that the generation of bi-directional narrow QFP wave trains is likely due to the periodic release of energy during a flare. This bi-directional wave train aligns with the findings of a study conducted by \cite{2016ApJ...823..150T}, which proposed the existence of a magnetic tuning fork phenomenon. According to their simulation, these wave trains can be spontaneously generated by the backflow pushing the arms of the magnetic tuning fork outward and compressing the magnetic field of the arms. This mechanism is analogous to the sound produced by an externally driven tuning fork. }

\xp{

For the broad QFP waves, its propagation direction has a certain angle with the local background magnetic field lines. As shown in Figure 4 in the article by \cite{2024ApJ...962...42H}, this angle is close to 90$^\circ$. We assume that the coronal waves, a fast mode waves, that propagate perpendicular to the mainly radially oriented magnetic fiedl lines, the measured speed can be written as

\begin{equation}
	v_f=\sqrt{v^2_A+c^2_s}
\end{equation}
, where $v_A=B/\sqrt{4\pi\mu nm_p}$ is the Alfve\'n speed, and $c_s\propto\sqrt{T}$ is the sound speed. In the quiet-Sun corona, the temperature is on the order of 1-2 MK, giving $c_s= 150-$\speed{210}. The magnetic field strength could be written as
\begin{equation}
B=\sqrt{(v^2_f-{c^2_s})(4\pi\mu nm_{p})}
\end{equation}
, where $m_p=1.64\times10^{-24} $ g is the proton mass, $\mu$ the mean molecular weight, and $n$ the total particle number density (taken as $\mu=0.6$, $n=1.92n_e$ according to \cite{1982soma.book.....P}, followed by \cite{2015LRSP...12....3W}, with $n_e$ as the electron density). Assuming the measured speed $v$ of the QFP wave trains equal to the fast-magnetosonic speed $v_f$, and adopting $n_e=2.8\times10^8$ cm$^{-3}$ (the average value of \cite{2021ApJ...921...61L}) for the quiet-Sun, the magnetic field strength of the quite-Sun region is roughly estimated as 4.0$\pm$0.23 G with the $v$ equal to 535 $\pm$ \speed{31} (a mean value of speeds along the S4-S18).

 \zxp{For the parallel propagation to the magnetic field, the fast-magnetosonic speed can be reduced as 
 
\begin{equation}
	v_f=v_A
\end{equation}
. Thus, the magnetic field strength can be estimated as 
\begin{equation}
B=v_f\sqrt{4\pi\rho}=v_f\sqrt{4\pi\mu nm_p}
\end{equation}
The magnetic field strength of coronal loops for southward and northward propagating wave trains respectively are 11.7$\pm$1.2\,G and 4.7$\pm$0.23\,G, with the $n_e=3.5\times10^8$ cm$^{-3}$ (taken from \cite{2022ApJ...941...59Z} ) and assume the measured speed $v$ equal to fast-magnetosonic speed $v_f$ (the speed of the southward and northward propagating narrow wave trains are 1417$\pm$\speed{145} and 565$\pm$\speed{28}, respectively). Notably, these computed magnetic field strength values should have significant uncertainties, which should be smaller than the actual value because the speeds utilized here are smaller than the actual speeds since the projection effect.
 }
 }

Similar to the report by \cite{2014ApJ...786..151S,2014ApJ...795..130S}, \cite{2023ApJ...959...71D} reported a Moreton wave confirmed by CHASE, where the accompanying EUV wave triggered a simultaneous horizontal and vertical oscillation of a quiescent filament. Unfortunately, we carefully checked the data from CHASE and other devices and found that the Moreton wave did not accompany the present event. This may be because the filament was not a highly inclined eruption, resulting in the excited large-scale broad EUV wave train being unable to effectively disturb the chromosphere \citep{2023ApJ...949L...8Z}. As the Sun enters the solar cycle 25, it is becoming more and more active \citep{2023SCPMA..6629631C}, and we look forward to discovering more Moreton waves, \zxp{possibly Moreton waves with} multiple wavefronts, combining the CHASE and other observation instruments.

In summary, the SDO data's excellent spatial and temporal resolution provides us with more details of the coronal waves. The origin of the broad and narrow QFP wave trains is still subtle. These wave trains are probably not dominated by a single excitation mechanism but are the result of a combination of excitation mechanisms discussed above. Thus, more observation with different wavebands is required to fully understand their exciting mechanism, evolution, and effect on the coronal plasma.

\section{Acknowledgments}
We would like to thank the anonymous referee, Prof. Dr. Bo Li, Prof. Dr. Ding Yuan, and Dr. Jialiang Hu for the many valuable suggestions and comments for improving the quality of this paper, and the teams of SDO, GOES, CHASE, SUTRI, and NVST for providing the excellent data. This work is supported by the Natural Science Foundation of China (12303062), the Natural Science Foundation for Youths of Sichuan Province (2023NSFSC1351), and the Project Supported by the Specialized Research Fund for State Key Laboratories. Y.D.S is supported by the Natural Science Foundation of China (12173083) and the Yunnan Science Foundation for Distinguished Young Scholars (202101AV070004). Co-author Ahmed Ahmed Ibrahim would like to thank the Researchers Supporting Project number (RSPD2024R993), King Saud University, Riyadh, Saudi Arabia. H.F.L is supported by Joint Funds of the National Natural Science Foundation of China (U1931116). We gratefully acknowledge ISSI-BJ for supporting the international team ``Magnetohydrodynamic wave trains as a tool for probing the solar corona''. Wavelet software was provided by C. Torrence and G. Compo, and is available at URL: \url{https://paos.colorado.edu/research/wavelets/}.


\vspace{5mm}
\end{CJK*}
\end{document}